\documentclass[prl,preprint,superscriptaddress,a4paper,showpacs]{revtex4}
\usepackage{amsmath,graphicx}
\usepackage{color}
\begin{document}

\title{Separating spin and charge transport in single wall carbon nanotubes}
\author{N. Tombros, S.J. van der Molen and B.J. van Wees}

\affiliation{Physics of Nanodevices, Materials Science Centre,
Rijksuniversiteit Groningen, Nijenborgh 4, 9747 AG Groningen, The
Netherlands}

\date{\today}
\begin{abstract}
\textbf{We demonstrate spin injection and detection in single wall
carbon nanotubes using a 4-terminal, non-local geometry. This
measurement geometry completely separates the charge and spin
circuits. Hence all spurious magnetoresistance effects are
eliminated and the measured signal is due to spin accumulation
only. Combining our results with a theoretical model, we deduce a
spin polarization at the contacts, $\alpha_F$, of approximately 25
$\%$. We show that the magnetoresistance changes measured in the
conventional two-terminal geometry are dominated by effects not
related to spin accumulation.}
\end{abstract}
 \maketitle

Single wall carbon nanotubes (SWNTs) behave as almost ideal
one-dimensional conductors, having a small diameter (typically a
nanometer) on the one hand, and a large scattering mean free path
on the other hand.\cite{Baughman} Additionally, it is expected
that electronic spin flip scattering in SWNTs is weak. This makes
them excellent candidates for spintronic devices, in which the
nanotubes are contacted by ferromagnetic leads. Despite the
promise that the combination of nanotubes and spintronics holds,
there have been no experiments so far that unequivocally
demonstrate spin accumulation in carbon nanotubes. In fact, all
experiments performed since the pioneering work of Tsukagoshi
\textit{et al.}\cite{Tsukagoshi}, have made use of the
conventional two-terminal spin valve geometry.
\cite{Tsukagoshi2,Kim,Chakraborty,Orgassa,Zhao,Sahoo,Jensen}
Unfortunately, in this geometry, it is difficult to separate spin
transport from other effects, such as Hall effects, anisotropic
magnetoresistance,\cite{Jedema,Alphenaar} and magneto-coulomb
effects\cite{Ono}. These may obscure and even mimic the spin
accumulation signal. With a four-terminal non-local spin valve
geometry,\cite{Jedema,Johnson,Jedema2} one is able to completely
separate the spin current path from the charge current path.
Hence, the signal measured is due to spin transport only. With
this technique we unambiguously demonstrate spin accumulation in
single wall carbon nanotubes.
\\

\begin{figure}[htb]
\begin{center}
\includegraphics[width=10cm]{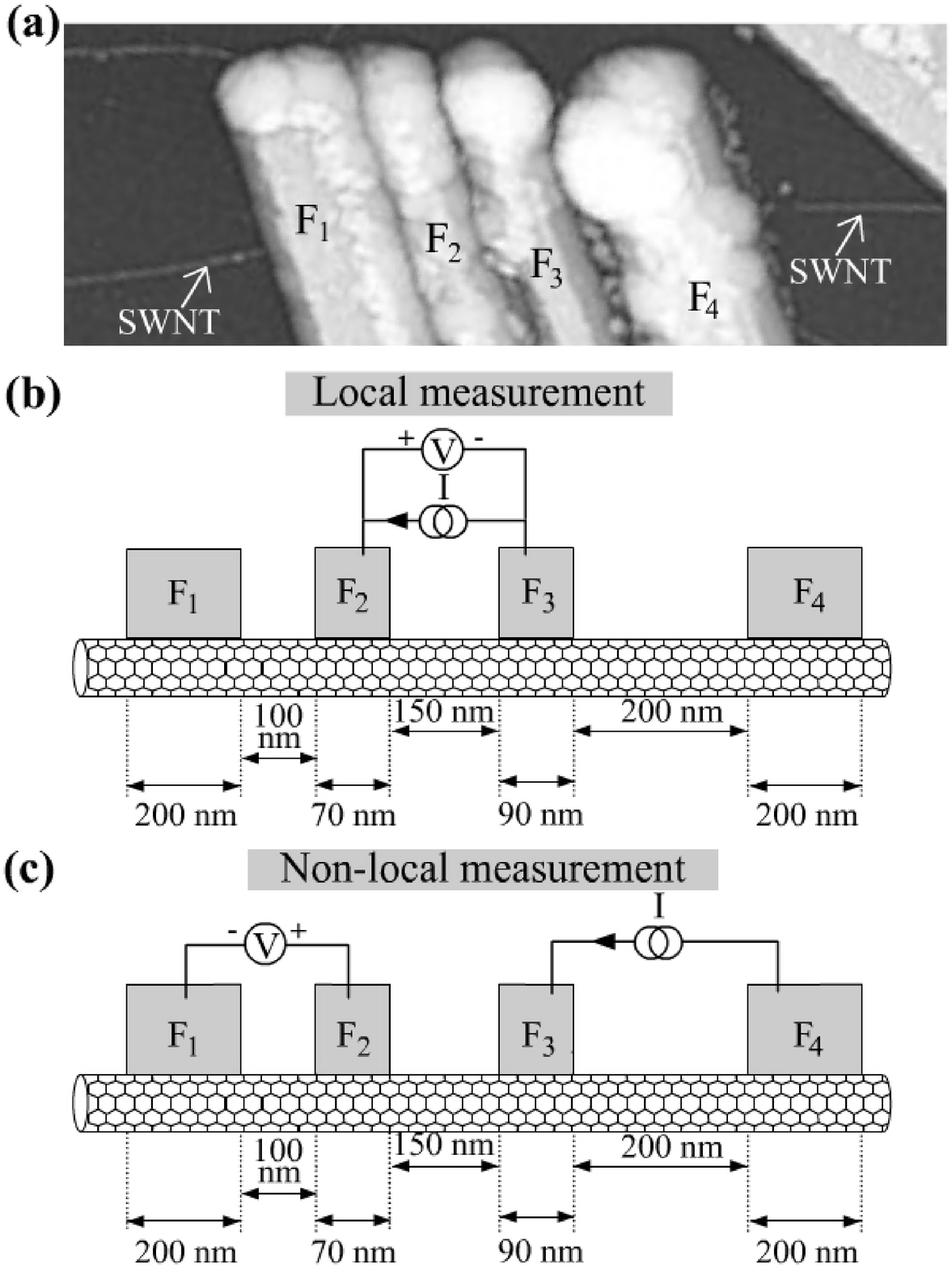}
\end{center}
\caption{A single wall carbon nanotube (d=3.4 $\pm$0.4 nm;
possibly it is a bundle containing a few nanotubes) contacted by
four ferromagnetic (cobalt) electrodes. (a) An AFM picture of the
device. Note that imperfect lift-off resulted in some PMMA residue
on top of the cobalt electrodes, this partially obscures the well
defined Co electrodes underneath.\cite{fn2} (b) Geometry of a
conventional spin valve (or 'local') measurement, in which
contacts F$_2$ and F$_3$ are used both to inject current and to
measure voltage. (c) The 'non-local' geometry. In this case the
voltage circuit (F$_1$-SWNT-F$_2$) is completely separated from
the current circuit (F$_3$-SWNT-F$_4$)}
\label{fig:device_local_nonlocal}
\end{figure}

To determine spin accumulation in the non-local geometry (see Fig.
\ref{fig:device_local_nonlocal}c)), one needs to contact a
metallic SWNT with four electrodes. At least two of these should
be ferromagnetic. They act as spin injector and spin detector,
respectively. For practical reasons, we make use of four
ferromagnetic contacts. These electrodes are narrow, but of
different widths to assure different switching fields $B_C$($B_C$
decreases with increasing width)\cite{Jedema,Jedema2}. Single wall
carbon nanotubes ($>$90$\%$ SWNT) \cite{SWNT} are dispersed in
HPLC grade chlorobenzene. We use the alternating current
dielectrophoresis technique \cite{Krupke} to deposit the SWNT's at
a predefined area on the substrate. An atomic force microscope
(AFM) in tapping mode is used to locate and characterize the
SWNT's on the SiO$_2$ surface. Conventional electron beam
lithography and e-beam evaporation (45 nm of Co at $4.0 \cdot
10^{-7}$ mbar) are used to define the contacts. To avoid damaging
the nanotube, no additional cleaning is done before deposition.

\begin{figure}[htb]
\begin{center}
\includegraphics[width=10cm]{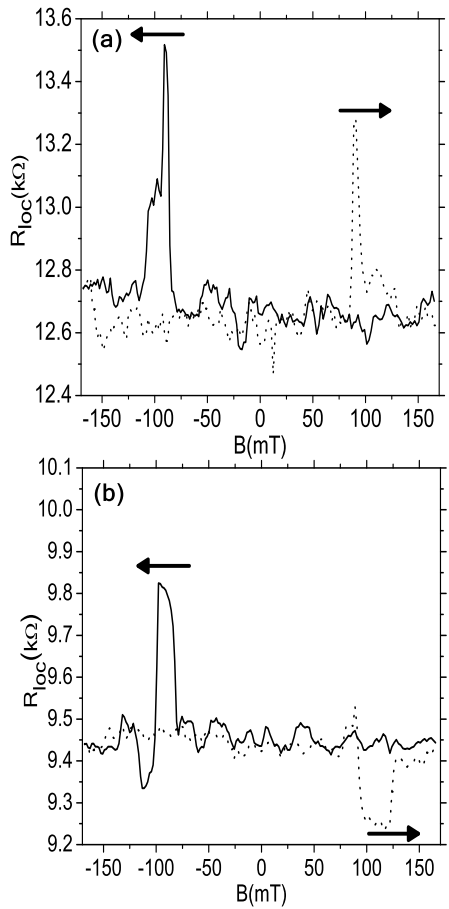}
\end{center}
\caption{Two-terminal spin valve measurements (F$_2$-SWNT-F$_3$, I
= 10nA) at 4.2 K (see Fig. \ref{fig:device_local_nonlocal}b). (a)
Upon sweeping the B-field, the 'local' resistance increases when
$|B|$ reaches $\approx 80 mT$. It falls back to its original value
when $|B|$ increases further. $\Delta$R/R has a maximum value of
$\approx 6 \%$. We observe significant substructure on top of the
resistance peaks. (b) A similar measurement on the same sample
(also at 4.2 K, but with a thermal cycling step in between). The
magnetoresistance trace is completely different from the traces in
a) and shows both positive and negative values for $\Delta$R/R.}
\label{fig:2ab_SV}
\end{figure}

\begin{figure}[htb]
\begin{center}
\includegraphics[width=7cm]{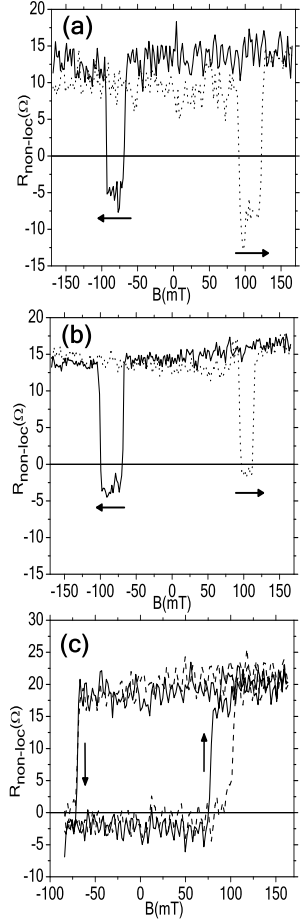}
\end{center}
\caption{Non-local measurement at T=4.2 K. The current path (from
F$_3$ to F$_4$) is separated from the voltage probes (F$_2$ to
F$_1$, see Fig. \ref{fig:device_local_nonlocal}c). The observed
resistance switching is due to spin accumulation and spin
transport in the single wall carbon nanotube. (a) Full magnetic
field scan with an a.c. current of 30 nA. We measure a negative
signal $R_{non-loc}$ when the spin injector, F$_3$, is
antiparallel to the spin detector, F$_2$. In this situation the
detector measures mainly the negative chemical potential of the
minority spin species. (b) Similar measurement to a), but now with
an a.c. current of 60 nA, resulting in a reduction of the noise
level. (c) The memory effect (I = 30nA), in which only the
magnetization of F$_3$ is switched.} \label{fig:3acmemory_NL}
\end{figure}

\begin{figure}[htb]
\begin{center}
\includegraphics[width=14cm]{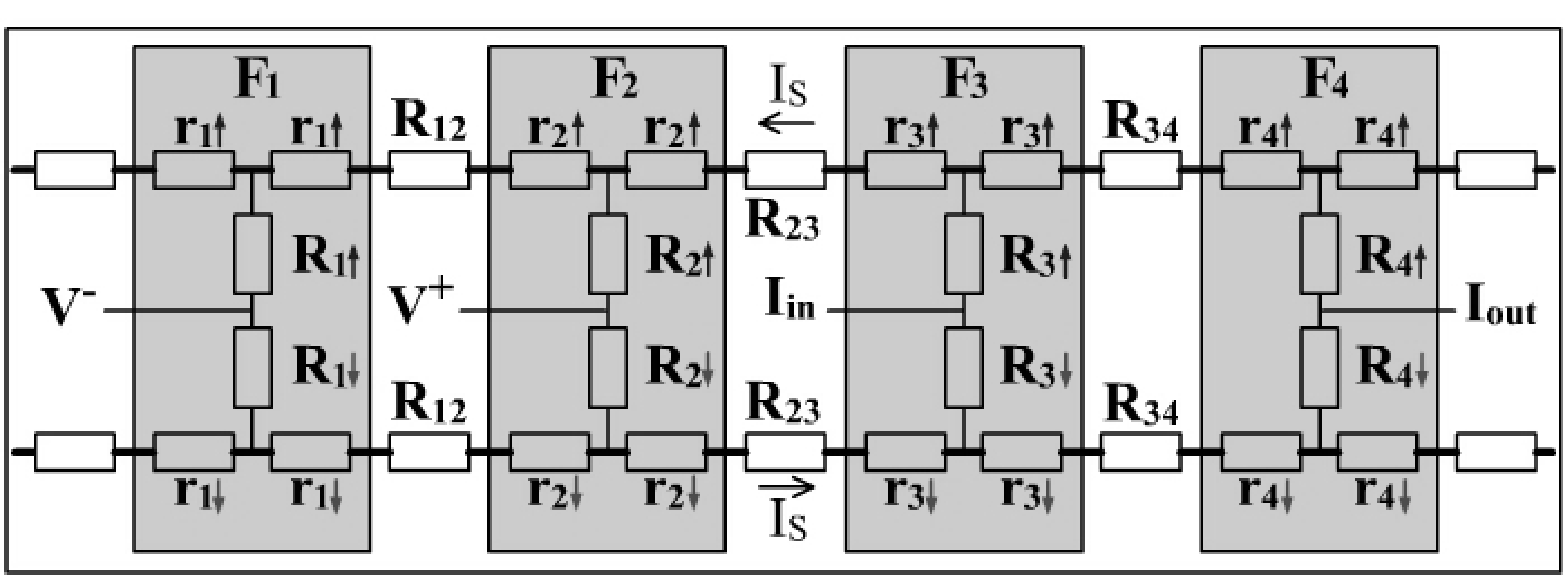}
\end{center}
\caption{A resistor model of our system (here, all four electrodes
are assumed magnetized in the 'up' direction). The upper half of
the resistor network corresponds to the spin up ($\uparrow$)
transport channel in the nanotube. The lower half to the spin down
($\downarrow$) channel. The resistance of the carbon nanotube
between the cobalt contacts F$_1$-F$_2$, F$_2$-F$_3$ and
F$_3$-F$_4$ is equal to $R_{12}/2$, $R_{23}/2$ and $R_{34}/2$,
respectively. The contact between the carbon nanotube and
ferromagnet F$_i$ (i=1,2,3,4) can be represented by a number of
spin-dependent resistances R$_{i,\eta}$ and r$_{i,\eta}$, where
$\eta = \uparrow,\downarrow$ denotes spin. Assuming spin up to be
the majority species, we have $R_{i,\uparrow}< R_{i,\downarrow}$
and $r_{i,\uparrow}< r_{i,\downarrow}$. Due to the spin-dependent
resistances in the current circuit (F$_3$ and F$_4$), the charge
current I produces a finite spin current I$_S$. Due to the
spin-dependent resistances in the voltage circuit (F$_1$ and
F$_2$), a non-zero voltage difference $V^+ - V^-$ consequently
develops, leading to a finite non-local resistance
$R_{non-loc}\equiv (V^+ - V^-)/I$.} \label{fig:ResistorModel}
\end{figure}

Although we regularly obtain low contact resistances
($\sim$k$\Omega$), the preparation of the device is not trivial,
as all the contacts have to be low ohmic. It is also crucial that
electron and spin transport can occur through the entire nanotube,
including the regions underneath the Co contacts. In Fig.
\ref{fig:device_local_nonlocal}, we depict our most successful
device. The two outermost electrodes, F$_1$ and F$_4$, have a
width of 200 nm. The two central electrodes, F$_2$ and F$_3$, have
a width of 70 nm and 90 nm, respectively. The nanotube itself has
a diameter of 3.4 $\pm$ 0.4 nm (we cannot exclude that it is a
bundle containing a few SWNT's). To measure the transport
properties of the nanotube, we make use of a standard a.c. lock-in
technique (maximum current: 60 nA). At 4.2 K, we find two-terminal
resistances of 28 k$\Omega$, 12.4 k$\Omega$ and 15 k$\Omega$
between contacts F$_1$-F$_2$, F$_2$-F$_3$ and F$_3$-F$_4$,
respectively. A four-terminal measurement (current from F$_1$ to
F$_4$; voltage between F$_2$ and F$_3$) gives a resistance of 10.3
k$\Omega$, equivalent to a conductance of $2.5\cdot e^2/h$. Since
this value is quite close to $4e^2/h$, the conductance of an ideal
ballistic nanotube, we are probing at least one metallic SWNT.
From the values above, we deduce the contact resistances between
the nanotube and electrodes F$_2$ and F$_3$. Comparing the
four-terminal resistance with the two-terminal measurement
(F$_2$-F$_3$), we get values around a k$\Omega$.
\\
Next, we investigate the two-terminal 'spin valve' effect between
contacts F$_2$ and F$_3$ (see Fig.
\ref{fig:device_local_nonlocal}b)). For this, we continuously
sweep the magnetic field back and forth between -165 mT and 165 mT
(at 4.2 K). Two characteristic traces are shown in Fig.
\ref{fig:2ab_SV}a). The behavior found is generally described as
follows. Let us start at $B=165 mT$, where F$_2$ and F$_3$ are
both magnetized parallel to the external field. When the field is
subsequently swept to negative values, F$_3$ (being the widest)
will flip magnetization as soon as the external field equals its
switching field. Consequently, the magnetizations of F$_2$ and
F$_3$ are now anti-parallel, leading to a resistance increase.
When the B-field gets more negative, also F$_2$ switches, so that
the magnetizations of both are parallel again. This leads to a
resistance decrease, back to the original value. A
magnetoresistance change of approximately 6 $\%$ is observed in
Fig. \ref{fig:2ab_SV}a). This is a considerable effect, comparable
to the values reported in Ref. \cite{Tsukagoshi} ($\leq 9 \%$).
\\
Although it appears that Fig.\ref{fig:2ab_SV}a) can be explained
as a result of spin transport only, we argue that this is not the
case. Figure \ref{fig:2ab_SV}b) shows an experiment performed on
the same sample in the exact same measurement geometry, at 4.2 K.
(There is a thermal cycling step in between Fig.
\ref{fig:2ab_SV}a) and b).) A completely different behavior is
observed. A predominantly negative, instead of positive,
magnetoresistance signal is now seen at positive B-fields. Similar
negative magnetoresistances have been observed in multiwall carbon
nanotubes.\cite{Chakraborty,Orgassa,Zhao} It is non-trivial to
explain these effects from spin transport only (they would require
a sign change in the polarization at an
electrode).\cite{Krompiewski} Another curiosity, often observed in
nanotubes (although not by us), is the fact that the
magnetoresistance increases before the external field has even
changed sign.\cite{Tsukagoshi,Alphenaar} The problem in the
interpretation lies in the fact that many other phenomena, not
related to spin, influence the magnetoresistance.
\cite{Jedema,Alphenaar,Ono}. Without extra knowledge these are
inseparable from spin accumulation in a two-terminal
experiment.\\
Fortunately, spin accumulation can be isolated from spurious
effects by adopting the non-local measurement geometry (see Fig.
\ref{fig:device_local_nonlocal}c)).\cite{Jedema,Johnson,Jedema2}
In such experiments, the \textit{charge} current path is
completely separated from the \textit{spin} current path. In our
case, this is done by attaching the current probes to F$_3$
(I$^+$) and F$_4$ (I$^-$) and the voltage probes to F$_2$ (V$^+$)
and F$_1$ (V$^-$), thus measuring the 'non-local'
magneto-resistance $R_{non-loc}\equiv (V^+ - V^-)/I$. In Fig.
\ref{fig:3acmemory_NL}, a) and b), we display two sets of
measurements. A clear and clean switching behavior is seen for all
traces. These results are similar to those obtained by Jedema
\textit{et al.} for Al wires.\cite{Jedema2} Characteristic is the
change of sign from positive (+15 $\Omega$) to negative (-5
$\Omega$) resistance values. This sign change can only happen if
the voltage probe F$_2$ ('detector') measures spin accumulation in
the SWNT system. In fact, when the voltage probe F$_2$ is parallel
to the spin 'injector' F$_3$, it probes the (positive)
electrochemical potential of the majority spin species (giving
positive non-local resistance). However, when its magnetization is
anti-parallel to that of F$_3$, it probes the (negative) chemical
potential of the minority spins. The change of sign thus assures
us that we are measuring spin accumulation (ruling out more
complicated current paths such as observed in multiwall
nanotubes)\cite{Bourlon}. We note that an important feature in
Fig. \ref{fig:3acmemory_NL} is the reduction of the noise
($\approx$3 $\Omega$), as compared to Fig. \ref{fig:2ab_SV}
($\approx$50 $\Omega$). This illustrates the insensitivity of
non-local measurements with respect to fluctuations in the overall
resistance.
\\
As an extra confirmation, we measure the so-called 'memory effect'
in the non-local geometry (Fig. \ref{fig:3acmemory_NL}c)). This
hysteresis effect is generated by allowing only one of the two
central electrodes to switch. We start at B=165 mT for which the
magnetizations of F$_2$ and F$_3$ are parallel and the non-local
resistance is positive. Subsequently, we decrease the magnitude of
the applied magnetic field to negative values until F$_3$ switches
at $\approx$ -70mT. Now F$_2$ and F$_3$ are anti-parallel and
$R_{non-loc}$ becomes negative. Next, we sweep to positive fields
again (F$_2$, F$_3$ are still anti-parallel) until at $\approx$70
mT electrode F$_3$ switches back. Thus we have returned to the
original situation. This demonstrates that the magnetization of
one individual electrode determines the sign of the measurement.
\\
Comparing Fig. \ref{fig:2ab_SV} with Fig. \ref{fig:3acmemory_NL},
an interesting observation can be made. Whereas in the
conventional spin valve measurement a magnetoresistance change
$\Delta R_{loc}\approx 700 \Omega$ is found, the 'non-local'
experiment yields $\Delta R_{non-loc}= 20 \Omega$, i.e., only 3
$\%$ of the 'local' value. This raises the question if the large
spin valve effect in Fig. \ref{fig:2ab_SV} originates from spin
accumulation or from spurious phenomena. To answer this, we model
the spin imbalance within the nanotube using a resistor network
(see Fig. \ref{fig:ResistorModel})\cite{model}. We assume the spin
flip length, $\lambda_{sf}$ to exceed all sample dimensions. The
(spin-independent) nanotube resistance between contacts $i$ and
$i+1$ is denoted by $R_{i,i+1}$. The contact resistance at each
electrode is split in two spin-dependent terms: $r_{i,\eta}$ and
$R_{i,\eta}$ (where $\eta=\uparrow,\downarrow$ denotes spin
direction). Both are calculated, assuming a contact conductivity
$\sigma_{\uparrow (\downarrow)} = \sigma_0 (1 \pm \alpha_F)/2$,
where $0 < \alpha_F < 1$ denotes the spin polarization. We have
measured the
 resistance and all possible combinations of two-,
three- and  four-probe resistances in the SWNT device. From this
we can determine the contact resistances between the nanotube and
contacts F$_2$ and F$_3$. The contact resistances between the
nanotube and contacts F$_1$ and F$_4$ cannot be precisely
determined and are assumed equal to those of F$_2$ and F$_3$. From
the model, the contact resistances, and the non-local traces, we
obtain a spin polarization $\alpha_F \approx 0.25$ \cite{error}.
We note that this is only a factor of two smaller than what is
ideally attainable. This indicates that the assumption of
$\lambda_{sf}$ being large in a carbon nanotube is justified. Now,
one can also calculate the expected 'local' resistance change,
giving $\Delta R_{loc} \approx 70 \Omega$.\cite{Jedema3}
Consequently, around 90 $\%$ of the resistance change in Fig.
\ref{fig:2ab_SV} cannot be attributed to spin accumulation. This
demonstrates how easily spin accumulation is masked by other
effects.\cite{Jedema,Jedema2} Interestingly, from the resistor
model, another phenomenon can be understood: the influence of
F$_1$ or F$_4$ on the non-local measurement is very small. The
reason is that the contribution due to these contacts is
attenuated by a factor of roughly $R_{i,\eta}/R_{i,i+1} \ll 1$. We
estimate that a magnetization change in the outer electrodes gives
a resistance change of around 1 $\Omega$, which lies within the
measurement noise.
\\
\\
Summarizing, we use a non-local measurement geometry to separate
spin transport from charge transport in a single wall carbon
nanotube contacted by ferromagnets. In this way, we unambiguously
demonstrate spin accumulation in a carbon nanotube device. Not
only does this work lead to a better understanding for future
spin-based nanotube applications, it also opens the road to more
sophisticated spin experiments on nanotubes (e.g. precession
measurements and/or determination of the spin flip length,
$\lambda_{sf}$, in carbon nanotubes).
\\

$\b{Acknowledgements}$

We thank Bernard Wolfs, Siemon Bakker, Anthony England, Edgar
Osorio, Marius Costache, Mihai Popinciuc and Steve Watts for
technical assistance and for useful discussions. This work was
financed by MSC$^{plus}$ and NWO (via a 'PIONIER' grant).

\newpage

\newpage
\newpage


\begin{thebibliography}{9}
\bibitem{error}
Given the assumptions of our model we estimate that the obtained
parameters are accurate within a factor of two
\end{thebibliography}
\end{document}